\documentclass[acmtog, nonacm]{acmart}

\acmSubmissionID{1261}

\usepackage{datenumber}
\usepackage{calc}
\usepackage[mmddyyyy]{datetime}
\newcounter{datetoday}
\newcounter{diffyears}
\newcounter{diffmonths}
\newcounter{diffdays}
\newcommand{\difftoday}[3]{%
      \setmydatenumber{datetoday}{\the\year}{\the\month}{\the\day}%
      \setmydatenumber{diffdays}{#1}{#2}{#3}%
      \addtocounter{diffdays}{-\thedatetoday}%
      \ifnum\value{diffdays}>0
        \def\diffbefore{}%
        \def\diffafter{left}%
      \else
        \def\diffbefore{}%
        \def\diffafter{ago}%
        \setcounter{diffdays}{-\value{diffdays}}%
      \fi
      \setcounter{diffyears}{\value{diffdays}/365}%
      \setcounter{diffdays}{\value{diffdays}-365*\value{diffyears}}%
      \setcounter{diffmonths}{\value{diffdays}/30}%
      \setcounter{diffdays}{\value{diffdays}-30*\value{diffmonths}}%
      \diffbefore
      \ifnum\value{diffyears}=0
      \else
        \ifnum\value{diffyears}>1
            \thediffyears\space years,
        \else
            \thediffyears\space year,
        \fi
      \fi
      \ifnum\value{diffmonths}=0
      \else
        \ifnum\value{diffmonths}>1
            \thediffmonths\space months
        \else
            \thediffmonths\space month
        \fi
      \fi
      \ifnum\value{diffdays}=0
      \else
        \ifnum\value{diffdays}>1
            \thediffdays\space days
        \else
            \thediffdays\space day
        \fi
      \fi
      \diffafter
}

\usepackage{microtype}

\usepackage{booktabs} 

\setcitestyle{square}

\usepackage{amsbsy}
\usepackage[mathscr]{euscript}
\usepackage{amsmath}
\usepackage{stackrel}
\usepackage{nicefrac}
\usepackage{kotex}

\usepackage{algpseudocode}
\usepackage[ruled]{algorithm2e} 

\SetAlFnt{\small}
\SetAlCapFnt{\small}
\SetAlCapNameFnt{\small}
\SetAlCapHSkip{0pt}
\IncMargin{-\parindent}

\usepackage{dblfloatfix}
\usepackage{float}

\usepackage{subcaption}

\begin{document}

\title{Hyperspectral Polarimetric BRDFs of Real-world Materials}

\author{Yunseong Moon}
\email{yunseong0518@postech.ac.kr}
\affiliation{%
  \institution{POSTECH}
  \country{South Korea}
}

\author{Ryota Maeda}
\email{maeda.ryota.elerac@gmail.com}
\affiliation{%
  \institution{POSTECH}
  \country{South Korea}
}
\affiliation{%
  \institution{University of Hyogo}
  \country{Japan}
}

\author{Suhyun Shin}
\email{}
\affiliation{%
  \institution{POSTECH}
  \country{South Korea}
}

\author{Inseung Hwang}
\email{ishwang@vclab.kaist.ac.kr}
\affiliation{%
  \institution{KAIST}
  \country{South Korea}
}

\author{Youngchan Kim}
\email{kyc618@postech.ac.kr}
\affiliation{%
  \institution{POSTECH}
  \country{South Korea}
}

\author{Min H. Kim}
\email{minhkim@kaist.ac.kr}
\affiliation{%
  \institution{KAIST}
  \country{South Korea}
}

\author{Seung-Hwan Baek}
\email{shwbaek@postech.ac.kr}
\affiliation{%
  \institution{POSTECH}
  \country{South Korea}
}

\authorsaddresses{}

\begin{abstract}
Acquiring bidirectional reflectance distribution functions (BRDFs) is essential for simulating light transport and analytically modeling material properties. 
Over the past two decades, numerous intensity-only BRDF datasets in the visible spectrum have been introduced, primarily for RGB image rendering applications. 
However, in scientific and engineering domains, there remains an unmet need to model light transport with polarization--a fundamental wave property of light--across hyperspectral bands.
To address this gap, we present the first hyperspectral-polarimetric BRDF (hpBRDF) dataset of real-world materials, spanning wavelengths from 414 to 950\,nm and densely sampled at 68 spectral bands. 
This dataset covers both the visible and near-infrared (NIR) spectra, enabling detailed material analysis and light reflection simulations that incorporate polarization at each narrow spectral band.
We develop an efficient hpBRDF acquisition system that captures high-dimensional hpBRDFs within a feasible acquisition time. 
Using this system, we demonstrate hyperspectral-polarimetric rendering using the acquired hpBRDFs.
To provide insights on hpBRDF, we analyze the hpBRDFs with respect to their dependencies on wavelength, polarization state, material type, and illumination/viewing geometry. 
Also, we propose compact representations through principal component analysis and implicit neural hpBRDF modeling.
Dataset is available on our project page\footnote{\url{https://yunseong0518.github.io/projects/hpBRDF/}}.
\end{abstract}

\begin{teaserfigure}
  \includegraphics[width=\linewidth]{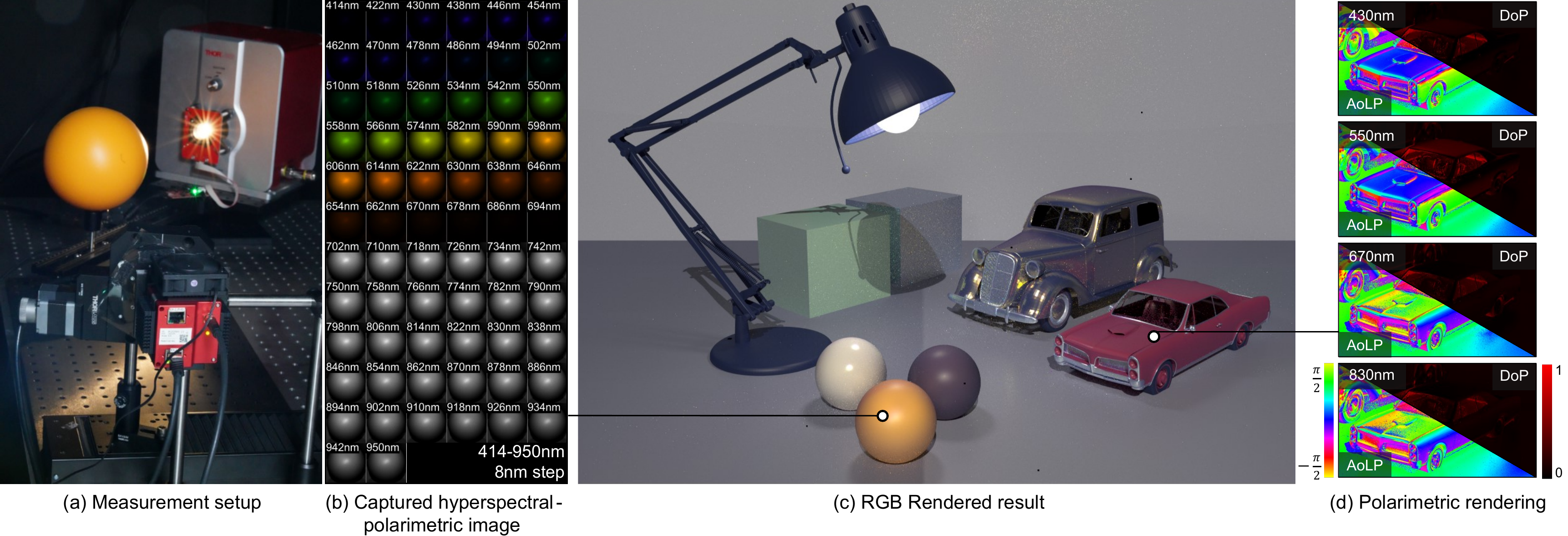}
  \vspace{-7mm}
  \caption{We present the first dataset of hyperspectral polarimetric BRDFs of real-world materials acquired by (a) our imaging setup, providing full Mueller-matrix BRDF measurements at (b) 68 spectral channels from visible to near-infrared (414–950\,nm). This enables rendering of not only (c) intensity but also (d)~polarization states from visible to NIR spectral bands.
  }
  \label{fig:teaser}
\end{teaserfigure}



\maketitle


\newcommand{\vect}[1]{\mathbf{#1}}
\newcommand{\mat}[1]{\mathbf{#1}}

\newcommand\note[1]{\textcolor{red}{#1}}

\newcommand{\psf}{\rho}

\newcommand{\argmin}[1]{\stackrel[\{ #1 \}]{}{\textrm{arg min}}}
\newcommand{\minimize}[1]{\stackrel[\{ #1 \}]{}{\textrm{minimize}}}

\section{Introduction}
\label{sec:intro}
Modeling how light reflects from materials is a fundamental problem across science and engineering, extending beyond computer graphics. 
The bidirectional reflectance distribution function (BRDF) models surface reflection as a 4D function over the 2D incident and 2D outgoing directions. 
BRDFs are central to physically based rendering for simulating light transport, and to inverse rendering for recovering geometry and material properties~\cite{pharr2023physically}.

Measured BRDF datasets have significantly advanced BRDF research over the past two decades. 
Early datasets focused on RGB intensity, facilitating realistic RGB rendering~\cite{Matusik:2003, marschner2000image}. 
With the increased importance of RGB-near-infrared (NIR) imaging and display applications, Dupuy and Jakob \shortcite{dupuy2018adaptive} expanded BRDF measurements from 360\,nm to 1,000\,nm, enabling rendering across visible and NIR spectra.

On the other hand, the emergence of polarimetric imaging has driven the development of polarimetric BRDF (pBRDF) datasets. 
Baek et al.~\shortcite{baek2020image} introduced a pBRDF dataset at five visible wavelengths, enabling polarimetric rendering at the five-selected spectra. 
However, five discrete wavelengths are insufficient to capture the spectral variability of polarization even within the visible spectrum.

Spectral and polarimetric approaches are also interconnected. 
Polarimetric inverse rendering leverages wavelength-dependent polarization cues to enhance shape reconstruction and reflectance decomposition~\cite{baek2018simultaneous, kondo2020accurate, hwang2022sparse, han2024nersp, li2024neisf, ichikawa2023fresnel}. 
Several studies have combined spectral-polarimetric modeling~\cite{huynh2013shape, kim2023neural, ha2024polarimetric}. 

Despite this progress, the lack of hyperspectral-polarimetric BRDF (hpBRDF) datasets limits the accuracy of modeling and simulation for applications that rely on spectral and polarimetric data.

To address this gap, we introduce a hpBRDF dataset, spanning 68 spectral bands from 414\,nm to 950\,nm, with a full width at half maximum (FWHM) of 10\,nm and a step size of 8\,nm. 
The dataset covers the full angular domain of surface reflection and is represented using Mueller matrices, enabling accurate modeling and simulation of arbitrary polarization states at each spectral band. 
This enables the rendering and analysis of complex wavelength-dependent polarization phenomena in real-world materials.

The primary challenge in acquiring an hpBRDF dataset is its high dimensionality, necessitating prolonged capture times using traditional scanning approaches. 
To efficiently acquire this high-dimensional data, we have developed an hpBRDF acquisition system integrating image-based BRDF acquisition~\cite{matusik2003data}, single-shot hyperspectral imaging, and broadband visible-to-NIR optical ellipsometry. 
Our system employs a hyperspectral light-field camera for single-shot visible-to-NIR hyperspectral imaging, significantly reducing acquisition times. 
We also implemented optical ellipsometry with accurate spectro-polarimetric calibration for broadband visible-to-NIR operation. 

Using our newly acquired real-world hpBRDF dataset comprising 14 materials, we demonstrate hyperspectral-polarimetric rendering. 
Furthermore, we analyze the dataset to explore dependencies on wavelength, polarization state, material characteristics, and illumination/viewing geometry. 
We also perform principal component analysis (PCA) and introduce an implicit neural representation of the hpBRDF, providing compact data representations. 
The neural approach additionally supports interpolation and continuous spectral-angular modeling.

In summary, our key contributions are:
\begin{itemize}
    \item A visible-to-NIR hpBRDF dataset for 14 real-world materials in Mueller-matrix form.
    \item An efficient hpBRDF acquisition system, including the acquisition procedure, calibration, and processing methods.
    \item An analysis of hpBRDF dependencies on wavelength, polarization state, and material characteristics, complemented by compact data representations using PCA and implicit neural representations.
\end{itemize}

\section{Related Work}
\label{sec:related}
\subsection{Acquisition Systems}
\paragraph{BRDF Acquisition}
BRDF measurements involve capturing reflected light intensity across combinations of incident and outgoing directions. 
Various optomechanical systems have been developed to facilitate these measurements, each balancing trade-offs in speed, accuracy, and flexibility. 
The gonio-reflectometer is a classical device for BRDF measurement~\cite{white1998reflectometer}. 
Conventional gonio-reflectometers measure light intensities in various light/view directions by rotating a light source and a sensor across a hemispherical range around a planar target. 
This approach has been extensively used for capturing BRDFs in both spectral~\cite{filip2014template, tsuchida2005development, li2006automated, dupuy2018adaptive} and polarization domains~\cite{kondo2020accurate}. 
However, achieving dense angular sampling with this approach is inherently time-consuming.
To overcome this limitation, an image-based BRDF measurement method has been proposed~\cite{marschner2000image, matusik2003data, ngan2005experimental}. 
This method captures convex objects, such as spheres or cylinders, instead of planar targets. 
A camera can capture dense angular combinations in a single shot, owing to the geometric variation of the objects, significantly reducing acquisition time. 
This strategy has been used for both spectral~\cite{kim2010acquisition} and polarization~\cite{baek2020image} measurements. 
In this work, we also adopt an image-based approach, utilizing a sphere object to efficiently acquire hpBRDF within a practical time frame.

\paragraph{Spectral Acquisition}
Early work on BRDF measurements often relied on a conventional RGB camera, which provided only limited spectral information. 
In contrast, multi- or hyperspectral imaging techniques acquire reflectance data in more than three bands. 
Spectrometers can measure thousands of spectral bands and have been integrated with gonio-reflectometers~\cite{dupuy2018adaptive}. 
However, the spectrometer only measures a single point and is unsuitable for image-based BRDF acquisition. 
To achieve spectral sampling in an image-based setup, previous approaches have employed multiple band-pass filters mounted on a rotating stage~\cite{tsuchida2005development, li2006automated, baek2020image} or a liquid crystal tunable filter~\cite{kim2010acquisition}. 
However, these time-multiplexing approaches require capturing additional images for each spectral band, substantially increasing measurement time. 
In contrast, our approach utilizes a hyperspectral light field camera capable of capturing 68 spectral bands in a single shot. 
This reduces the total acquisition time, making hpBRDF measurement feasible.

\begin{figure*}[t]
  \includegraphics[width=\linewidth]{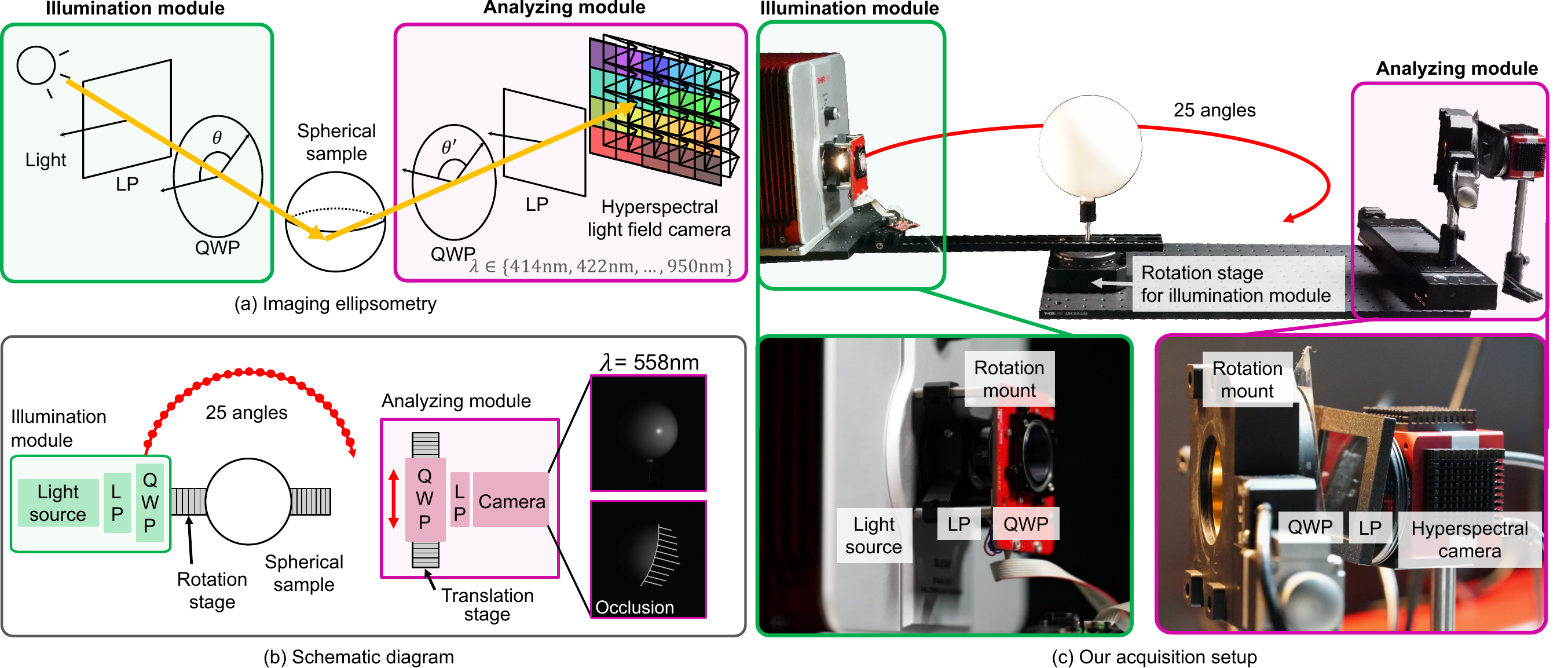}
  \caption{\textbf{hpBRDF acquisition system.} 
  (a) Our system consists of an illumination module and an analyzing module. 
  The illumination module emits broadband polarized light, which is reflected at a sample sphere. 
  The analyzing module captures the reflected light using a hyperspectral light field camera, which captures different spectral images from multiple viewpoints. 
  Both the illumination and analyzing modules have achromatic LP and QWP to generate and analyze the polarization state. 
  (b) The illumination module is rotated around the sample using a motorized rotation stage. 
  To address occlusion, we translate the QWP and acquire two separate image sets. 
  (c) Photographs of our acquisition setup. 
  }
    \label{fig:setup}
\end{figure*}

\paragraph{Polarimetric Acquisition}
Polarimetric reflectance can be described by a Mueller matrix, which characterizes how a surface changes the polarization state of incident light. 
Such measurements are commonly performed using an ellipsometer~\cite{collett2005field}, which systematically modulates the polarization states on both the light source and detector sides.
Kondo et al.~\shortcite{kondo2020accurate} measured a pBRDF using rotating linear polarizers, thereby obtaining a 3$\times$3 Mueller matrix limited to linear polarization components. 
Baek et al.~\shortcite{baek2020image} constructed a pBRDF dataset that captures the full 4$\times$4 Mueller matrix, encompassing both linear and circular polarization effects. 
Their measurement employed a dual-rotating-retarder (DRR) method~\cite{azzam1978photopolarimetric}, which rotates quarter-wave plates (QWPs) to sample a comprehensive range of polarization states. 
Following this approach, our method also adopts the DRR technique to obtain a complete 4$\times$4 Mueller matrix for visible-to-NIR wavelengths. 

\subsection{BRDF Datasets}
The MERL dataset~\cite{matusik2003data} and the UTIA dataset~\cite{filip2014template} each contain BRDF measurements of approximately one hundred materials, but only at RGB intensity. 
Dupuy and Jakob~\shortcite{dupuy2018adaptive} introduced a hyperspectral BRDF dataset spanning the visible to NIR domain. 
Baek et al.~\shortcite{baek2020image} introduced a pBRDF dataset that includes polarization properties at five discrete wavelengths in the visible spectrum. 
Our hpBRDF dataset offers both hyperspectral and polarimetric measurements. 

\subsection{Applications of Spectral-polarimetric Imaging}
Our hpBRDF dataset provides an opportunity for exploring the relationships between spectral and polarimetric visual information. Parametric pBRDF models~\cite{priest2000polarimetric, hyde2009geometrical, baek2018simultaneous, kondo2020accurate, hwang2022sparse, ichikawa2023fresnel, ha2024polarimetric} are fundamental to not only forward rendering, but also inverse rendering for geometry and appearance estimation~\cite{baek2018simultaneous, dave2022pandora, kondo2020accurate, hwang2022sparse, ha2024polarimetric, han2024nersp, li2024neisf}. 
Yet, most existing pBRDF models neglect spectral dependencies on pBRDF. 
On the other hand, incorporating spectral-polarimetric dependencies has been found to be useful for accurately capturing various optical phenomena. 
For example, wavelength-dependent variations in the refractive index have been leveraged for shape recovery from a spectral-polarimetric image~\cite{huynh2013shape} and for glass segmentation in RGB-polarization images~\cite{mei2022glass}. 
Moreover, large-scale spectral-polarimetric datasets have begun to uncover complex interrelationships between spectral and polarization characteristics in natural images~\cite{jeon2023spectral}. 
Our hpBRDF dataset paves the way for developing more physically grounded hpBRDF models and spectro-polarimetric imaging applications.

\section{\texorpdfstring{\MakeLowercase{hp}}~BRDF Imaging System}
\label{sec:acquisition}
To efficiently acquire hpBRDFs of real-world materials, we design a hyperspectral-polarimetric reflectance imaging system that combines spectral and polarization sensitivity with angular sampling efficiency.
Figure~\ref{fig:setup} illustrates our hpBRDF imaging system. 
It captures the hpBRDF of a spherical object, represented as a Mueller matrix $\mathbf{M}$, within a practical acquisition time. 
To this end, we extend image-based BRDF imaging~\cite{matusik2003data} with single-shot hyperspectral imaging and visible-to-NIR optical ellipsometry.




\subsection{Polarized VIS-NIR Illumination}
The illumination module of our imaging system uses a stabilized quartz tungsten-halogen lamp (\textit{Thorlabs SLS301L}), which produces a consistent visible-to-NIR spectral output denoted by $L^\lambda$. 
Since the emitted light is unpolarized, its Stokes vector is:
$\mathbf{s}^\lambda = [L^\lambda, 0, 0, 0]^\intercal$.
The unpolarized light passes through two polarization-modulating components: an ultra-broadband linear polarizer (LP, \textit{Thorlabs WP25M-UB})  characterized by Mueller matrix $\mathbf{L}$, and an achromatic quarter-wave plate (QWP, \textit{Thorlabs AQWP10M-580}) mounted on a high-speed motorized rotation stage (\textit{Thorlabs ELL14K}). 
By rotating the QWP to an angle $\theta$, multiple polarization states can be generated. 
The resulting Stokes vector of the light exiting the illumination module at wavelength $\lambda$ is given by:
\begin{equation}
\label{eq:light}
\mathbf{s}^\lambda_\text{emitted}(\theta) = \mathbf{R}^\lambda(\theta)  \mathbf{L}  \mathbf{s}^\lambda,
\end{equation}
where $\mathbf{R}^\lambda(\theta)$ is the Mueller matrix of the QWP rotated to angle $\theta$. 
This configuration enables rapid and flexible control over the polarization state of broadband light across the visible to near-infrared spectrum.
The entire illumination module is mounted on a heavy-duty motorized rotation stage (\textit{Thorlabs HDR50/M}), which rotates around the target sample as shown in Figure~\ref{fig:setup}(b). 

\begin{figure*}[t]
    \includegraphics[width=\linewidth]{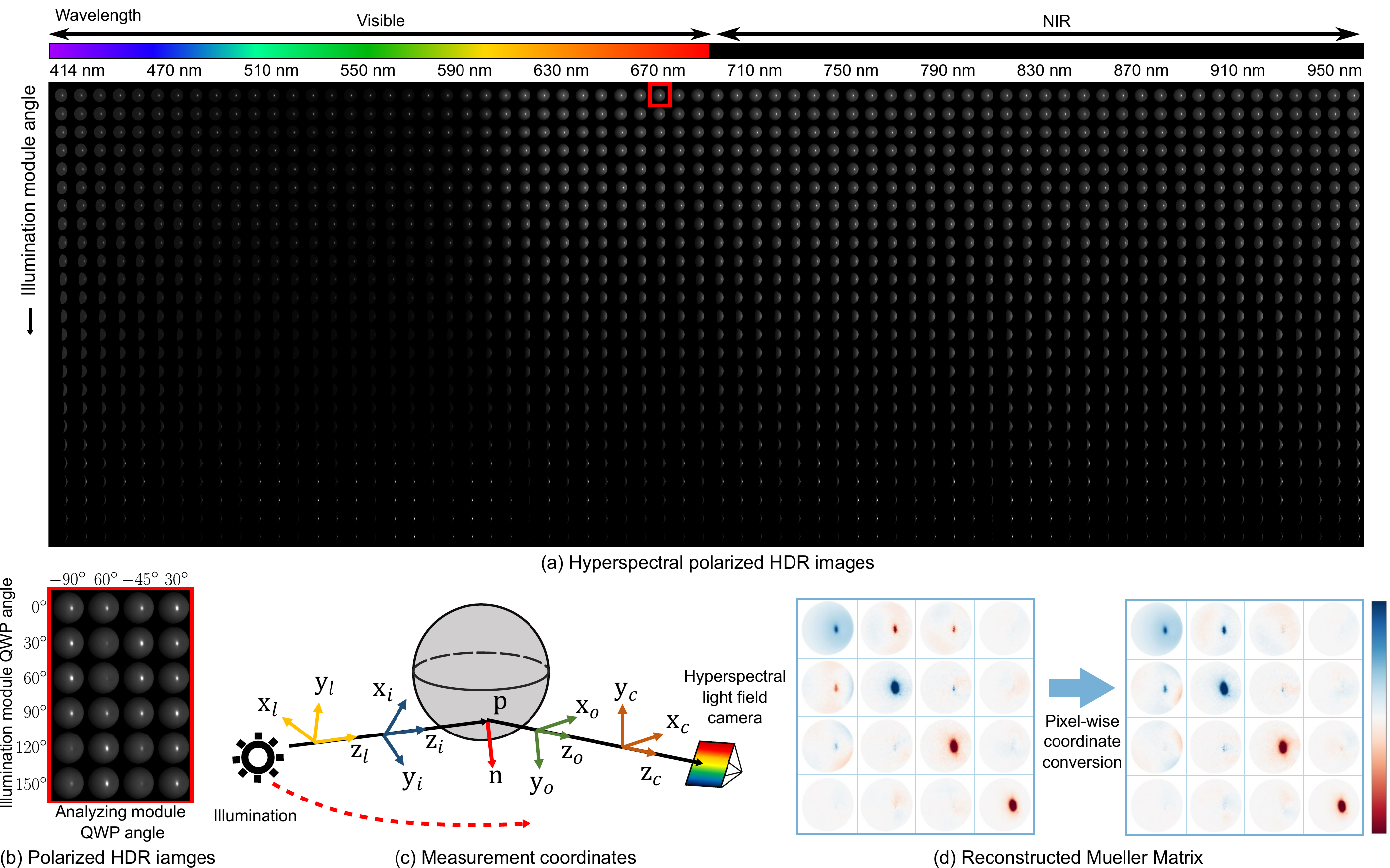}
    \caption{
    \textbf{hpBRDF acquisition.} 
    (a) Hyperspectral HDR images captured under different rotation angles of the illumination module. 
    (b) Captured images with varying QWP angles corresponding to a specific configuration marked by the red rectangle in (a). 
    (c) Given the coordinate systems of the illumination and analyzing modules, ($\mathbf{x}_l, \mathbf{y}_l, \mathbf{z}_l$), ($\mathbf{x}_c, \mathbf{y}_c, \mathbf{z}_c$), respectively, the hpBRDF coordinate is defined with respect to ($\mathbf{x}_i, \mathbf{y}_i, \mathbf{z}_i$) and ($\mathbf{x}_o, \mathbf{y}_o, \mathbf{z}_o$) that dependent on the TBN (tangent, bitangent, normal) space of the surface. 
    (d)~The reconstructed Mueller matrix is transformed to the hpBRDF coordinate. 
    }
    \label{fig:fig6}
        \vspace{-3mm}
\end{figure*}

\subsection{Reflection on Spherical Samples}
We capture homogeneous spherical samples as in previous image-based BRDF imaging~\cite{matusik2003data} as shown in Figure~\ref{fig:setup}(c). 
The continuous variation of surface normals across the sphere, combined with changing incident angles from the rotating light source, enables dense sampling over a wide angular domain.
For a surface point on the sphere, the incident light direction is denoted by $\omega_i$ and the direction of the reflected light by $\omega_o$. 
The hpBRDF at that point is defined as a Mueller matrix $\mathbf{M}(\omega_i, \omega_o)$, describing the polarization transformation of incident light. 
The reflected Stokes vector is:
\begin{equation}
\label{eq:reflection}
\mathbf{s}^\lambda_\text{reflected}(\theta,\omega_o) = \mathbf{M}(\omega_i, \omega_o)\mathbf{s}^\lambda_\text{incident}(\theta,\omega_i).
\end{equation}
As shown in Figure~\ref{fig:fig6}(c), we apply a coordinate transformation to align the emitted-light Stokes vector and the incident-light Stokes vector as
$\mathbf{s}^\lambda_\text{incident}(\theta,\omega_i)=\mathbf{C}_{\text{e}\rightarrow\text{i}}\mathbf{s}^\lambda_\text{emitted}(\theta)$.


\subsection{Hyperspectral Polarimetric Imaging}
\label{AnalyzingModule}
\paragraph{Visible-to-NIR Polarimetric Modulation}
Our analyzing module measures the Stokes vector $\mathbf{s}^\lambda_\text{reflected}(\theta, \omega_o)$ of the reflected light, as shown in Figure~\ref{fig:setup}(a). 
The light first passes through a 2-inch achromatic polymer QWP (\textit{Edmund Optics WP140HE}), mounted on a motorized rotation stage (\textit{Thorlabs HDR50/M}) and rotated to angle $\theta'$. 
It then passes through a broadband 2-inch linear polarizer (\textit{Thorlabs WP50L-UB}) mounted on a custom 3D-printed holder.
To mitigate ghosting artifacts caused by the internal reflections between the QWP and the LP, both components are tilted by approximately 15$^{\circ}$ with respect to the optical axis. The selected tilt angles are within the specified operational range of each component. 
The resulting polarization-modulated Stokes vector is:
\begin{equation}
\label{eq:capture}
\mathbf{s}^\lambda_\text{analyzed}(\theta', \omega_o) = \mathbf{L} \mathbf{R}^\lambda(\theta’) \mathbf{s}^\lambda_\text{captured}(\theta, \omega_o),
\end{equation}
where, as shown in Figure~\ref{fig:fig6}(c), coordinate conversion is applied to the reflected-light Stokes vector: 
$\mathbf{s}^\lambda_\text{captured}(\theta,\omega_i)=\mathbf{C}_{\text{r}\rightarrow\text{c}}\mathbf{s}^\lambda_\text{reflected}(\theta)$.

\paragraph{Single-shot Hyperspectral Imaging}
The light is then captured by a hyperspectral light-field camera (\textit{Cubert Ultris X20}), which uses an array of narrow bandpass filters integrated into its microlens array to enable single-shot acquisition of spectral data at each different viewpoint, as shown in Figure~{\ref{fig:setup}(a)}. 
This configuration minimizes acquisition time--critical for practical hpBRDF measurement. The hyperspectral camera provides a spatial resolution of 410$\times$410 pixels with 164 spectral bands by default. 
We select 68 channels at 8\,nm intervals from 414\,nm to 950\,nm to balance spectral resolution and data size, considering FWHM of the camera is 10\,nm.
The recorded intensity at wavelength $\lambda$ under the QWP angles of the illumination module and analyzing module $\theta$ and $\theta’$, respectively, is:
\begin{equation}
\label{eq:intensity}
f(\lambda, \theta, \theta') = [\mathbf{s}^\lambda_\text{analyzed}(\theta', \omega_o)]_0,
\end{equation}
where $[]_0$ extracts the first entry of the Stokes vector, representing the intensity of light.
Note that the outgoing direction $\omega_o$ varies per wavelength due to the multi-view geometry of the light-field camera. 
Because each spectral image corresponds to a micro-lens view direction, spatial misalignment between spectral bands is inherent. 
This necessitates geometric calibration across the spectral bands to correctly associate spatial and angular information in the hyperspectral-polarimetric image domain.

\paragraph{Occlusion Handling}
The 2-inch QWP in the analyzer module limits the effective field of view of the hyperspectral light-field camera. 
To overcome this, we perform two acquisitions with the QWP positioned at distinct lateral offsets, using a linear translation stage (\textit{Thorlabs NRT150}) as shown in Figure~\ref{fig:setup}(b). 
In the figure, we show the raw captured image for two different QWP positions for wavelength 558\,nm. For each spectral channel, we select an occlusion-free mask to ensure complete Mueller matrix reconstruction. 

\subsection{Calibration}
We perform geometric, radiometric, and polarimetric calibration of our hpBRDF imaging system. 
Geometrically, our hyperspectral light-field camera is modeled as an array of 68 multi-view cameras, each capturing distinct narrow spectral bands. 
We calibrate their intrinsic and extrinsic parameters using multiple checkerboard captures.
We also estimate the positions of the sample sphere via sphere fitting, and accurately determine light-source positions by minimizing the differences between simulated and captured specular highlights. 
Radiometrically, we calibrate the camera response by imaging a Spectralon reference with known diffuse reflectance across visible-to-NIR wavelengths, facilitating stable and linear HDR hyperspectral imaging. 
Polarimetrically, we precisely align LPs and QWPs using a polarimeter, allowing controlled polarization states through rotations of QWPs in both illumination and analyzing modules. 
For more details, please refer to the Supplementary Document.

\section{\texorpdfstring{\MakeLowercase{hp}}~BRDF Dataset}
\label{sec:hpbrdf}

\begin{figure}[t]
    \includegraphics[width=\linewidth]{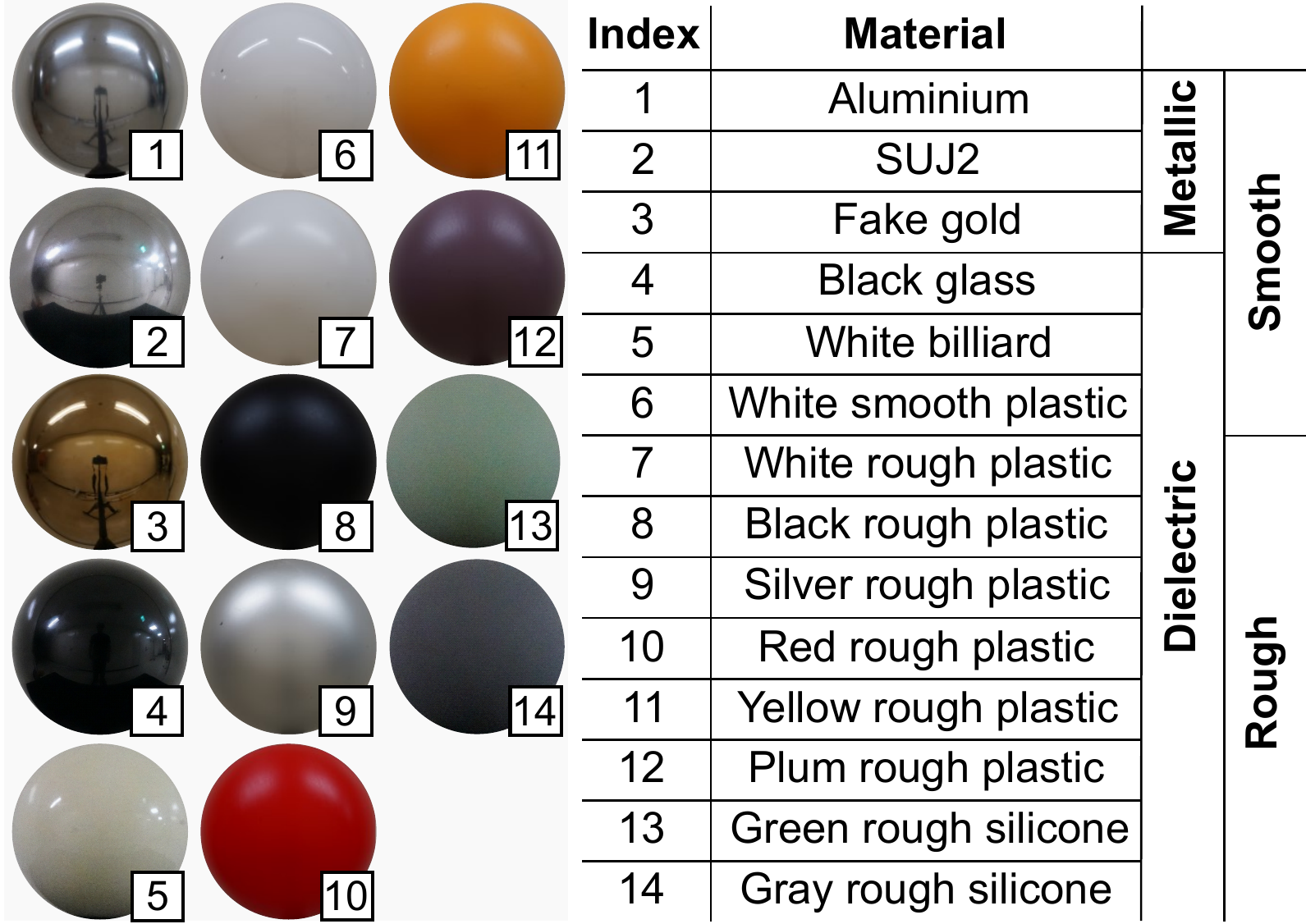}
    \caption{\textbf{Material samples.} Our dataset provides the measured hpBRDFs of 14 real-world spherical samples.}
    \label{fig:material_sample}
    \vspace{-5mm}
\end{figure}


Using our hpBRDF imaging system, we have acquired the hpBRDFs of 14 real-world materials as shown in Figure~\ref{fig:material_sample}.
We represent hpBRDF $\mathbf{M}$ in a 4D tabulation form: $\mathbf{P}(\lambda, \phi_d, \theta_d, \theta_h) \in \mathbb{R}^{4\times4}$, where we use Rusinkiewicz coordinate~\shortcite{rusinkiewicz1998new} of $\phi_d, \theta_d, \theta_h$ for the incoming and outgoing light directions discretized into 361, 91 and 91 bins respectively. 
For the spectral axis, the hpBRDF table has 68 bins from 414\,nm to 950\,nm with 8\,nm interval. For each table element at the index of $\lambda, \phi_d, \theta_d, \theta_h$, the table stores a $4\times4$ Mueller matrix in a single-precision float, leading to 13\,GB per single hpBRDF table.

\paragraph{Image Acquisition} 
\label{ImageAcquisitionProcess}
To acquire hpBRDF table, we first capture images covering all combinations of system parameters, including illumination-module angle, analyzing-module QWP translation, QWP-angle configurations, and exposure levels. 
Specifically, for angular sampling, the illumination module is rotated through 25 discrete angles, ranging from 40$^\circ$ to 160$^\circ$ at intervals of 5$^\circ$. 
At each illumination angle, two positions of the analyzing-module QWP are used using a translation stage to mitigate occlusion caused by the rotation mount, as shown in Figure~\ref{fig:setup}. 
Ellipsometric measurements require rotating QWPs in both the illumination and analyzing modules to predefined angular configurations: specifically, angles $\Theta \in \{30^\circ, -45^\circ, 60^\circ, -90^\circ\}$ for the illumination module, and angles $\Theta' \in \{0^\circ, 30^\circ, 60^\circ, 90^\circ, 120^\circ, 150^\circ\}$ for the analyzing module. 
For each QWP and illumination configuration, we capture an HDR image by merging multiple exposure-bracketed images. The number of exposures ranges from 3 to 11, depending on the reflectance properties of the material.
Consequently, the total number of hyperspectral images acquired per material ranges between 3,744 and 12,480.
The total acquisition time per material ranges from 13 to 35 hours.

\begin{figure}[t]
    \includegraphics[width=\linewidth]{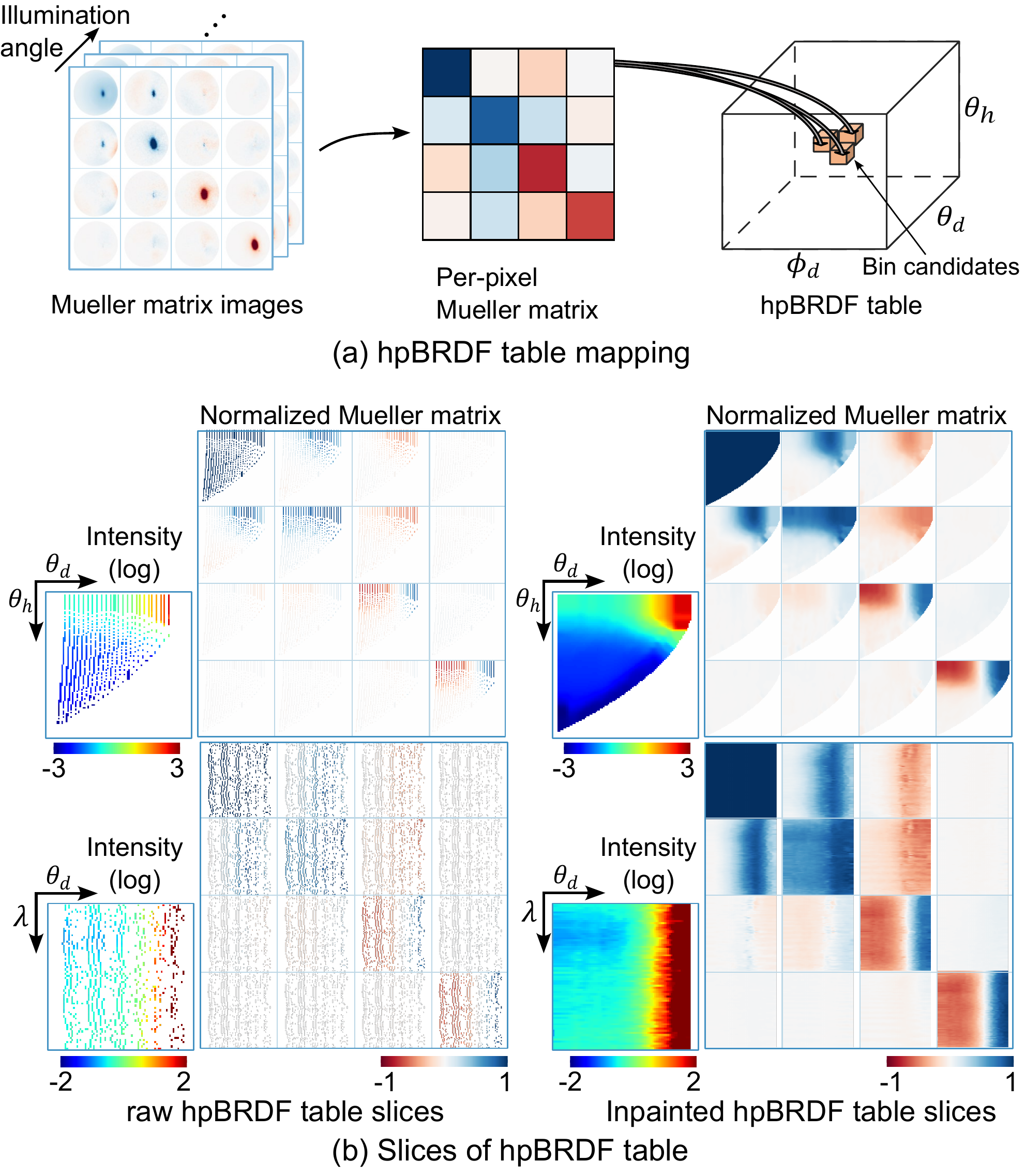}
    \vspace{-5mm}
    \caption{\textbf{hpBRDF table construction} 
    (a) We map per-pixel Mueller matrices to the neighbour bins in the hpBRDF table. 
    (b) Representative slices of the hpBRDF table for raw and inpainted versions. 
    }
    \label{fig:hpbrdf_table_construction}
    \vspace{-5mm}
\end{figure}

\paragraph{hpBRDF Reconstruction}
We then reconstruct hpBRDF $\mathbf{M}$ for each pixel, illumination angle, and wavelength channel. 
Using the image formation models of Equations~\eqref{eq:light}, \eqref{eq:reflection}, \eqref{eq:capture}, \eqref{eq:intensity}, this is formulated as a least-squares problem:
\begin{equation}\label{eq:measure_mm}
        \underset{\mathbf{M}}{\mathrm{minimize}}\sum_{\theta, \theta' } \left(f(\lambda,\theta,\theta')-[\mathbf{L} \mathbf{R}^\lambda(\theta’) \mathbf{C}_{\text{r}\rightarrow\text{c}}\mathbf{M}\mathbf{C}_{\text{e}\rightarrow\text{i}} \mathbf{R}^\lambda(\theta)  \mathbf{L}  \mathbf{s}^\lambda]_0\right)^2.
\end{equation}
Figure~\ref{fig:fig6}(d) shows the reconstructed Mueller matrix image. 

\paragraph{hpBRDF Table Construction}
To construct the hpBRDF table, we calculate per-pixel Rusinkiewicz coordinates for each Mueller matrix. 
As shown in Figure~\ref{fig:hpbrdf_table_construction}(a), we map a single pixel's Mueller matrix to the target hpBRDF table bin including the neighbor bins. 
Due to mechanical constraints in our imaging system, some entries are missing in the hpBRDF table especially for low $\theta_d$. We perform numerical inpainting of the missing entries. 
We use 3D Gaussian convolution in the $(\phi_d,\theta_d,\theta_h)$ space for each Mueller matrix element at each wavelength for inpainting.
Figure~\ref{fig:hpbrdf_table_construction}(b) shows the slices of the constructed hpBRDF table before and after applying the inpainting.
For constructing the hpBRDF table, it takes about 10 hours per material.

\section{Results}
\label{sec:results}


\begin{figure*}[t]
  \includegraphics[width=\linewidth]{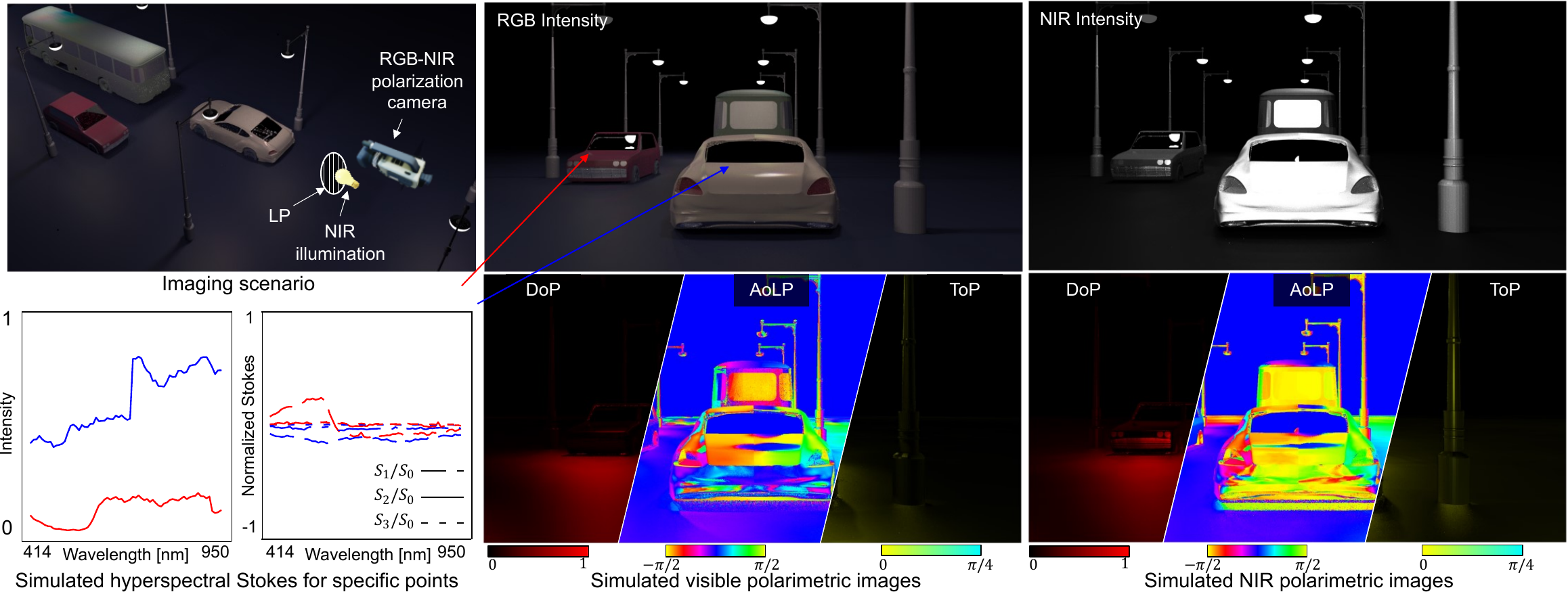}
  \caption{\textbf{Hyperspectral polarimetric rendering.} We simulated a driving scenario under nighttime illumination using our hpBRDF dataset. The scene includes multiple vehicles and halogen-based street lights. We use the imaging setup with an RGB-NIR polarization camera and an active linearly polarized NIR LED illumination, showing different polarization characteristics for the visible and NIR spectrum.}
  \vspace{-2mm}
  \label{fig:rendering_result}
\end{figure*}


\subsection{Validation}
We perform three validation experiments.
First, we verify the accuracy of our reconstructed Mueller matrices by comparing them with ground-truth measurements for air, linear polarizer, and quarter-wave plate as shown in Figure~\ref{fig:validation}(a). Results demonstrate close alignment with the ground truth. 
Second, we assess the physical validity of the reconstructed Mueller matrices. 
A physically valid Mueller matrix maps any physically admissible Stokes vector to another admissible Stokes vector. 
A Stokes vector $\mathbf{s} = [s_0, s_1, s_2, s_3]^\intercal$ is considered admissible if it satisfies: $s_0 \geq 0, \quad s_0^2 - (s_1^2 + s_2^2 + s_3^2) \geq 0$.
We use Givens–Kostinski's method~\shortcite{givens1993simple} to verify this condition for three selected materials from our raw hpBRDF dataset. 
The results indicate that 94.50\% of the reconstructed Mueller matrices satisfy the physical validity criteria across all hpBRDF bins.
A complete analysis of the physical validity results for all 14 materials are provided in the supplementary document.
Third, we qualitatively evaluate the fidelity of our measurements by comparing the hpBRDF data with similar visual appearances named ``white billiard" provided by Baek et al.~\shortcite{baek2020image}. 
{Figure~\ref{fig:validation}(b)} shows qualitative agreement between our dataset and the reference measurements at the five wavelengths. 


\subsection{Rendering Polarimetric Visible-NIR Images}
Using our hpBRDF dataset on Mitsuba~3~\cite{jakob2022mitsuba3} enables the simulation of polarized light transport for arbitrary scenes across the visible-NIR spectrum. 
Since our hpBRDF table follows the same format as Baek's pBRDF table~\shortcite{baek2020image}, this allows for rendering with minimal code modifications.
Figure~\ref{fig:rendering_result} shows an example of a nighttime driving scenario, exhibiting different polarization characteristics of the RGB and NIR spectra due to the use of polarized NIR illumination. 
Figure~\ref{fig:teaser} shows another example of hyperspectral polarimetric rendering. 


\subsection{hpBRDF Analysis}
\label{sec:albedo_polarimetric}

To analyze the polarization characteristics of our hpBRDF, we employ the Lu-Chipman decomposition~\shortcite{lu1996interpretation}, which factorizes each Mueller matrix into three physically interpretable matrices: a depolarization matrix, a retardance matrix, and a diattenuation matrix. We can also derive four scalar quantities: diattenuation, polarizance, polarization preservation, and retardance. For detailed mathematical formulations and definitions, refer to the Supplementary document.

\paragraph{Wavelength Dependency}
We analyze the wavelength-dependent behavior of hpBRDF. 
Figure~\ref{fig:img_based_analysis} shows diattenuation and preservation images of three plastic spheres with distinct colors: white, red, and yellow. 
The white plastic sphere exhibits uniformly high spectral intensity across the whole wavelength range, with minimal variation. 
In contrast, the red and yellow plastic spheres show pronounced spectral intensity changes within the visible wavelength range. 
Notably, preservation tends to increase at wavelengths with lower reflectance, whereas diattenuation remains relatively stable despite changes in spectral intensity.


\paragraph{Surface Roughness}
Figure~\ref{fig:roughness} shows that smooth surfaces exhibit significantly stronger polarization-preserving behavior than rough surfaces consistently over the entire measured spectrum.
This indicates that the polarization information of illumination is not preserved for rough surfaces, regardless of the light spectrum, potentially due to multiple scattering between microfacets.


\paragraph{Dielectric vs. Metallic} \label{subsec:metallicity}
Figure~\ref{fig:dielec_metal} shows the comparison of the dielectric and metallic material using retardance Mueller matrix images $\mathbf{M}_R$. While both materials exhibit a similar diagonal component, clear differences appear in non-diagonal components, particularly $\mathbf{M}_{R,13}$, $\mathbf{M}_{R,23}$, $\mathbf{M}_{R,31}$, $\mathbf{M}_{R,32}$. 
These components show sign inversions between metallic and dielectric materials, indicating distinct phase shift behaviors in reflection since the retardance matrix represents the phase changes. 



\subsection{hpBRDF Representations}

\begin{figure*}
    \centering
  \includegraphics[width=\linewidth]{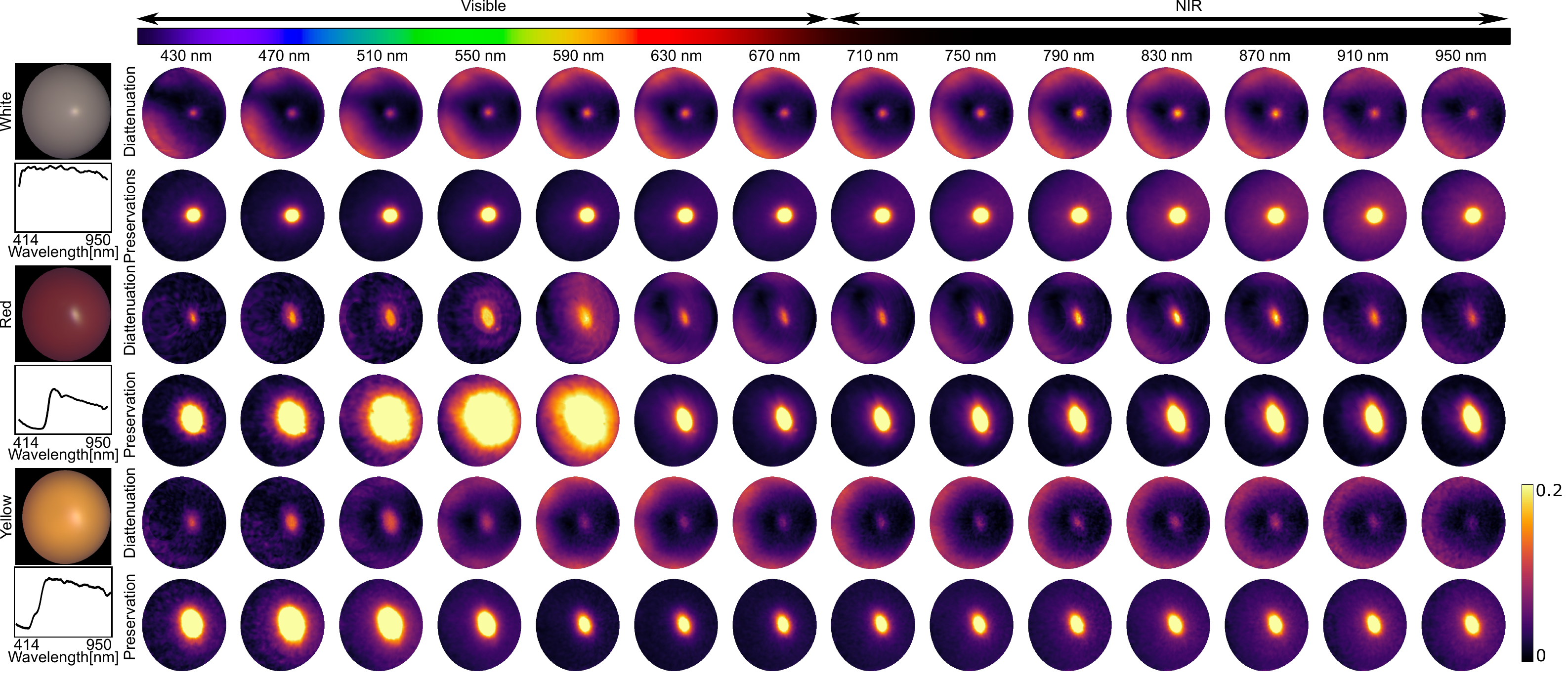}
  \caption{\textbf{Wavelength dependency of hpBRDF.} We visualize three colored rough plastic samples (white, red, yellow) and their diattenuation and polarization preservation across wavelengths. The result shows wavelength-dependent characteristics for each material. 
  }
  \label{fig:img_based_analysis}

  \vfill
    \vspace{2mm}
    \begin{minipage}{0.49\linewidth}
  \includegraphics[width=\columnwidth, angle=0]{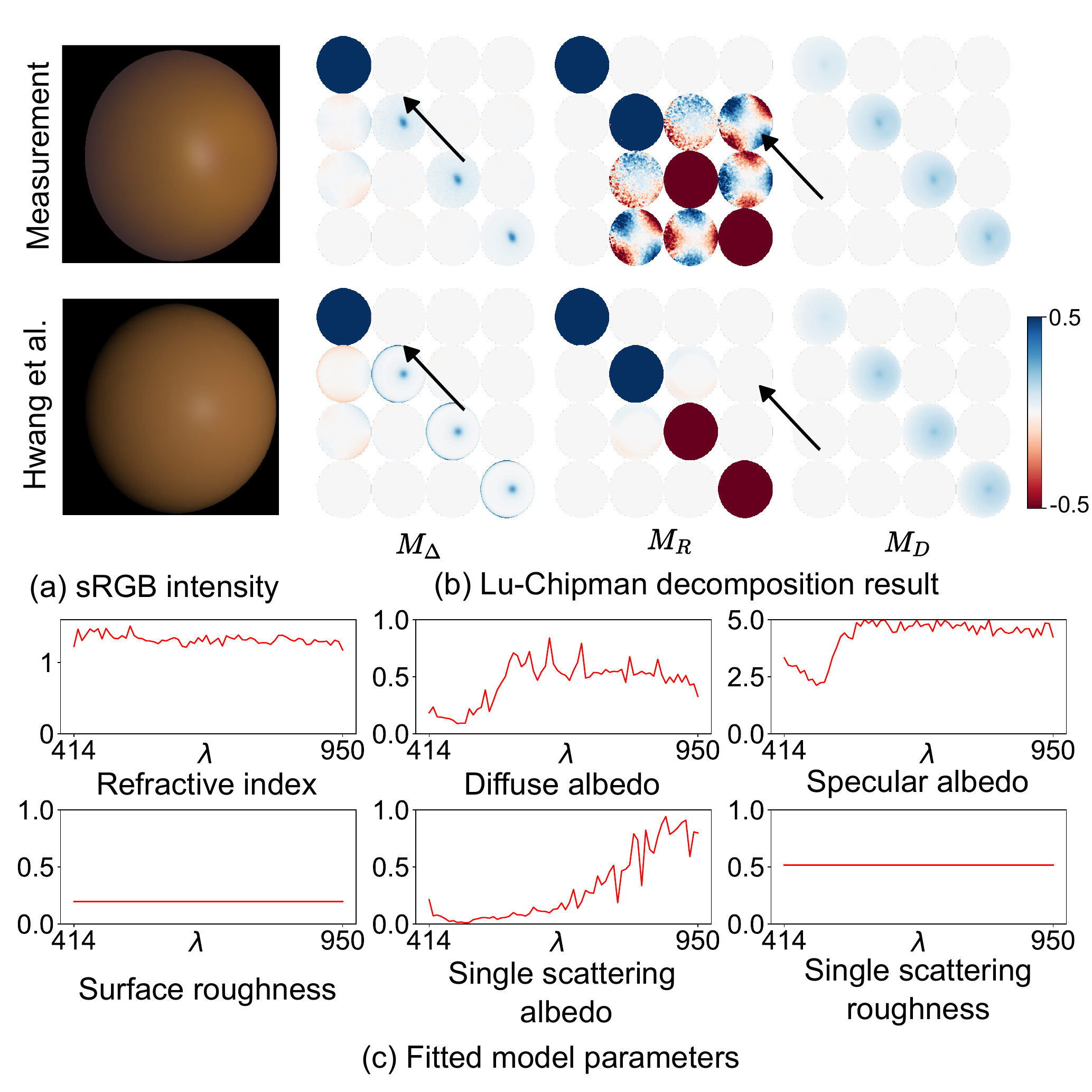}
    \caption{
    \textbf{Evaluation of analytical pBRDF model on our hpBRDF.} We fit analytic parametric pBRDF model~\cite{hwang2022sparse} to our yellow rough plastic hpBRDF table. We show the (a) rendered sRGB intensity and (b) Lu-Chipman decomposition result of specific wavelength (550\,nm) using our hpBRDF table and the fitted model. In the decomposition matrix $\mathbf{M}_\Delta$, we observe overestimation near the edges of the sphere. In the retardance matrix $\mathbf{M}_R$, the analytical model fails to represent the non-diagonal component patterns. (c) Fitted model parameters.
    }
    \label{fig:analytic_pbrdf}
    \end{minipage}
    \hspace{1.5mm}
    \centering
    \begin{minipage}{0.49\linewidth}
        \includegraphics[width=\columnwidth]{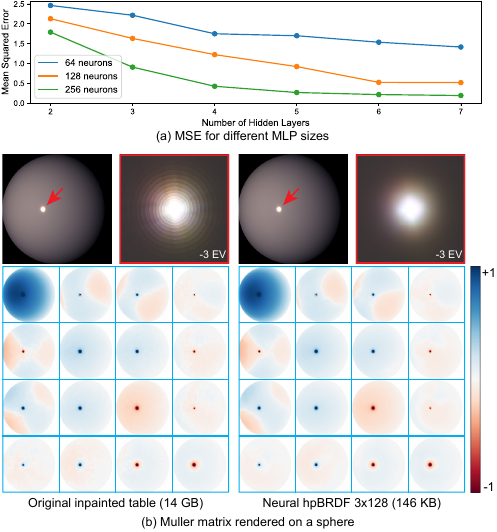}
        \caption{\textbf{Implicit neural representation for hpBRDF.} (a) Mean squared error (MSE) of the predicted Mueller matrices for "white smooth plastic", evaluated across different MLP configurations. 
        (b) Visualization of the original inpainted table and the neural hpBRDF corresponds to incident angles of $30^{\circ}$ and a wavelength of $\lambda = 670$\,nm. The neural hpBRDF closely matches the original Mueller matrix with continuous interpolation, while significantly reducing storage requirements. The rendered results are visualized with gamma correction ($\gamma = 2.2$).
        }
        \label{fig:neural_hpbrdf}
    \end{minipage}
\end{figure*}

\paragraph{Implicit Neural Representation}
We introduce an implicit neural representation for hpBRDF that enables continuous interpolation and extremely compact storage of hpBRDF table. 
Whereas recent neural BRDF models map incident and outgoing directions to RGB radiance~\cite{rainer2019neural, sztrajman2021neural, zeltner2024real, dou2024real}, our formulation extends the input domain with wavelength $\lambda$~\cite{kim2023neural, jeon2023spectral} and predicts the full $4{\times}4$ Mueller matrix, thereby jointly modeling spectral and polarization variation. 
Figure~\ref{fig:neural_hpbrdf} shows the trained result with (a) different network sizes and (b) visualization of the trained hpBRDF. 
The predicted Mueller matrices closely resemble the original inpainted table, well representing the whole hpBRDF. 
In the rendering with a table, nearest-neighbor lookups introduce discretization artifacts, resulting in noticeable discontinuities at specular highlights. In contrast, the neural hpBRDF provides a continuous prediction of the Mueller matrix across both directional and spectral domains, enabling smooth rendering.
Furthermore, the neural hpBRDF achieves a substantial reduction in storage size. For instance, a network with four hidden layers of 256 neurons in a single-precision floating-point format occupies 146~kB--about $10^{5}$ times smaller than the original 13~GB tabulation--while preserving visual fidelity.

\begin{figure*}
    \begin{minipage}{.49\linewidth}
    \includegraphics[width=\linewidth]{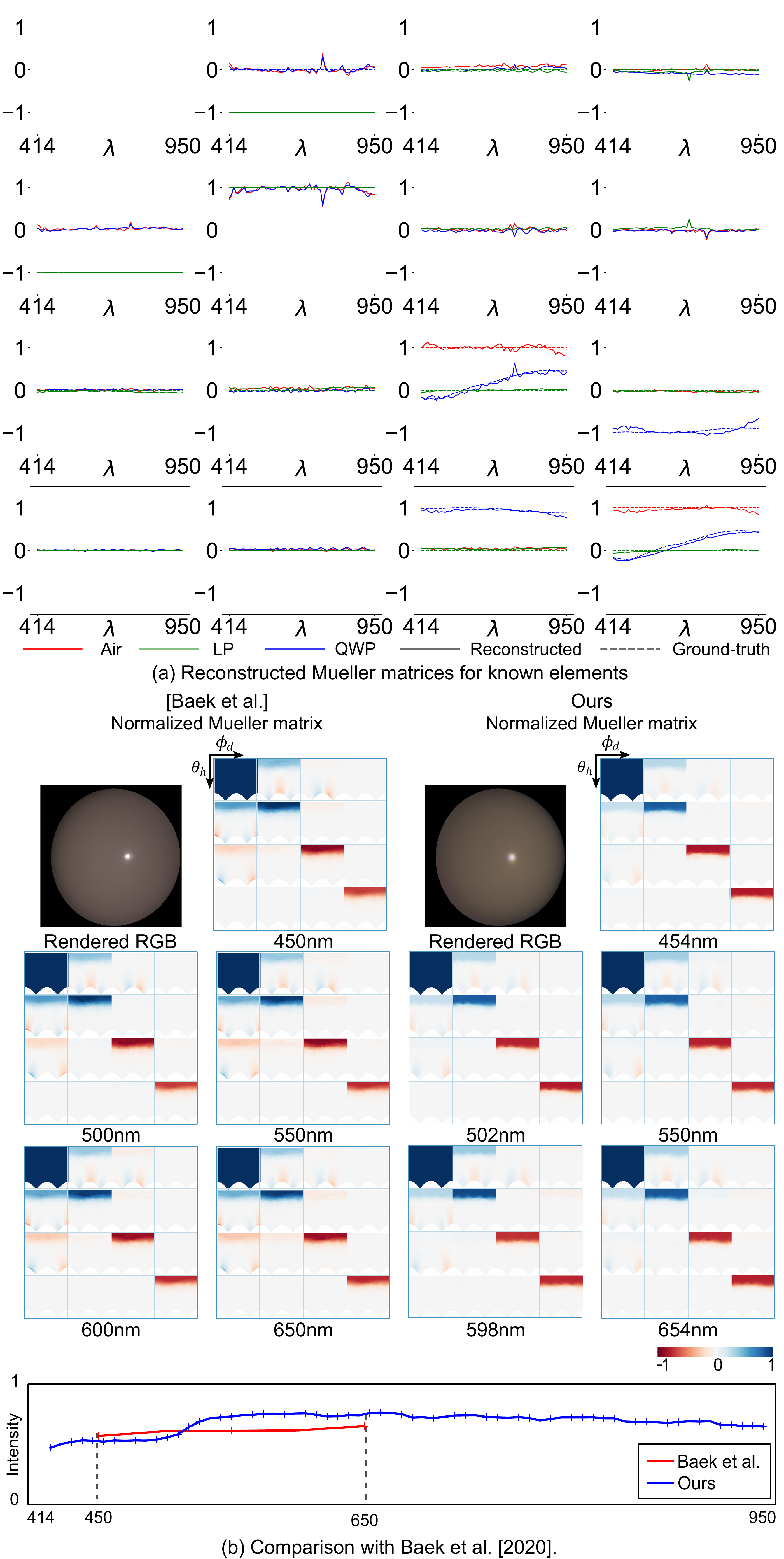}
    \caption{
    \textbf{Validation.} 
    (a) We validate our Mueller matrix reconstruction pipeline by capturing elements with known Mueller matrices. For QWP, the captured QWP has wavelength dependent retardance, so we calculated ground truth with known retardance. Reconstructed Mueller matrices are closely similar to the ground truth across the entire wavelength range. 
    (b) We compare our hpBRDF table with the existing pBRDF table~\shortcite{baek2020image}, using the ``white billiard" where it has a similar visual appearance. The pBRDF tables from both datasets show similar Mueller matrix, qualitatively validating the accuracy of our measurements. 
    Note, our dataset offers denser and broader spectral sampling, in contrast to the five discrete wavelengths in \cite{baek2020image}.
    }
    \label{fig:validation}
    \end{minipage}
    \hfill
    \begin{minipage}{.49\linewidth}
    \includegraphics[width=\linewidth]{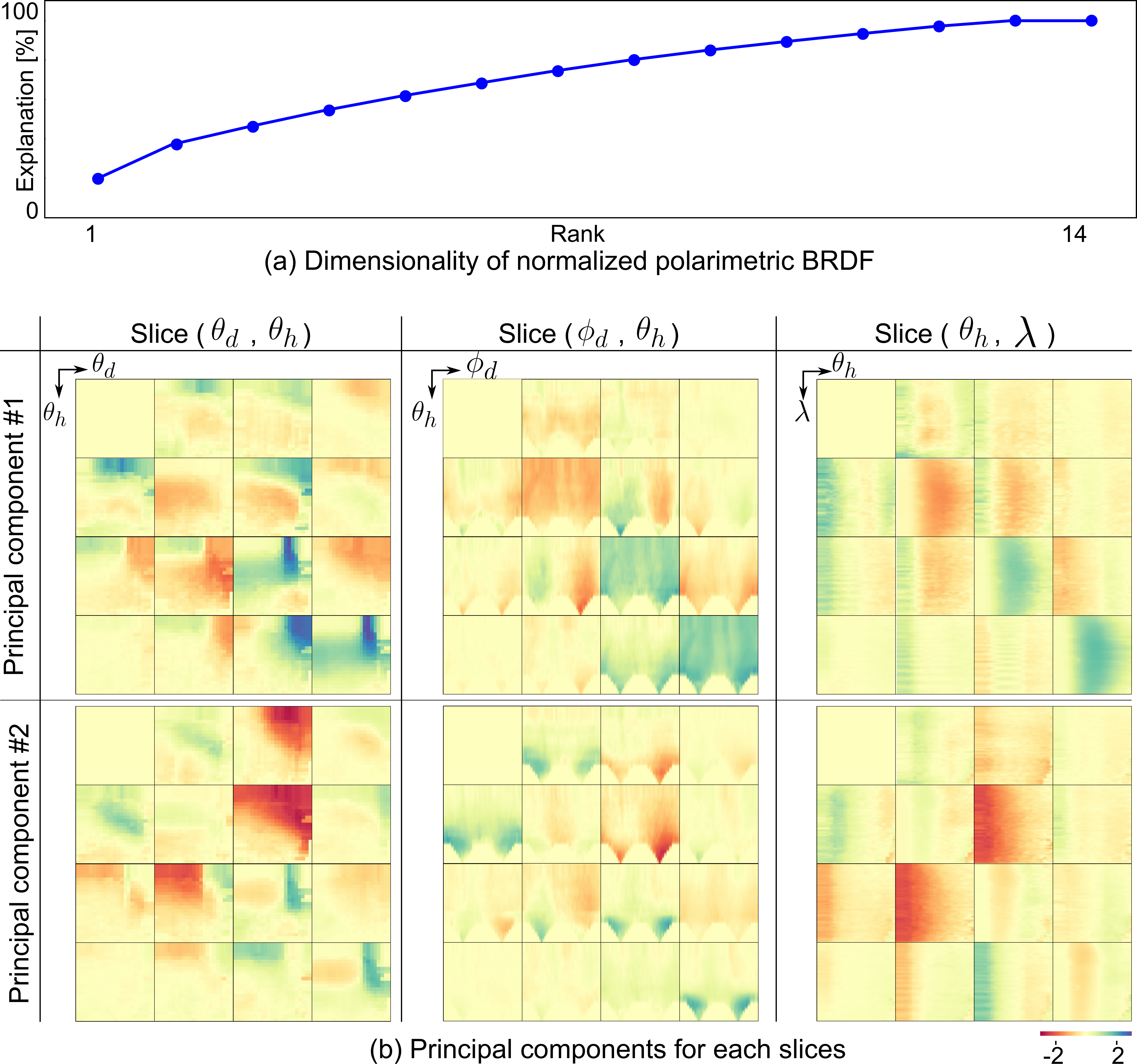}
    \caption{\textbf{PCA analysis of hpBRDFs.} (a) The dimensionality of hpBRDFs using PCA. (b) Estimated first two principal components.
    }
    \label{fig:pca_main}
    \includegraphics[width=\linewidth]{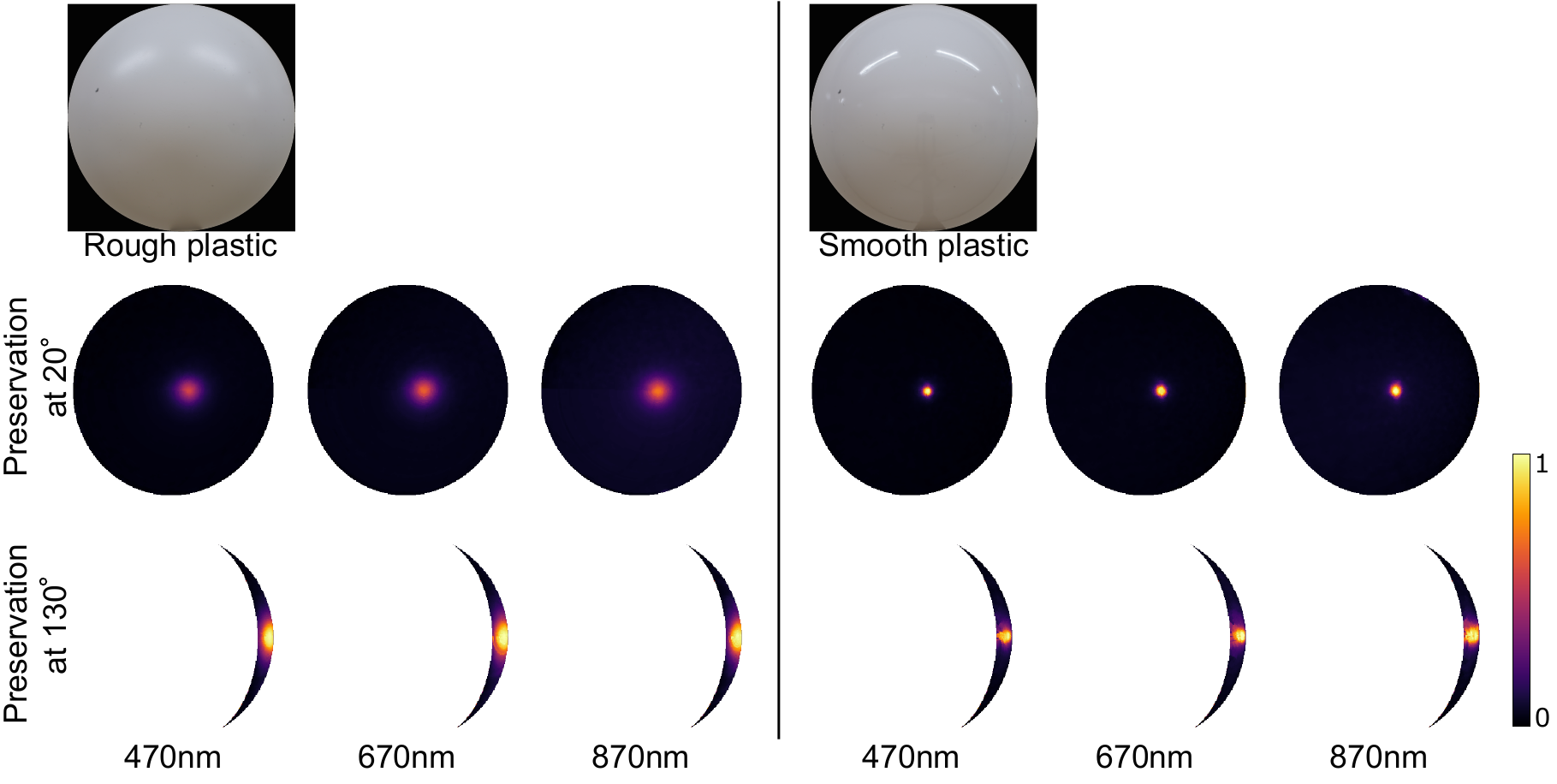}
    \vspace{-5mm}
    \caption{\textbf{Surface roughness and hpBRDF.} 
    For smooth material, the preservation value remains consistently close to 1 in specular highlight region, regardless of $\theta_d$. In contrast, the preservation value of rough material shows a sharp transition around a specific $\theta_d$ in specular highlight region. At high $\theta_d$ values, preservation value approaches 1, similar to smooth material, while at low $\theta_d$, it stabilizes a lower preservation value. 
    }
    \label{fig:roughness}
    \includegraphics[width=\linewidth]{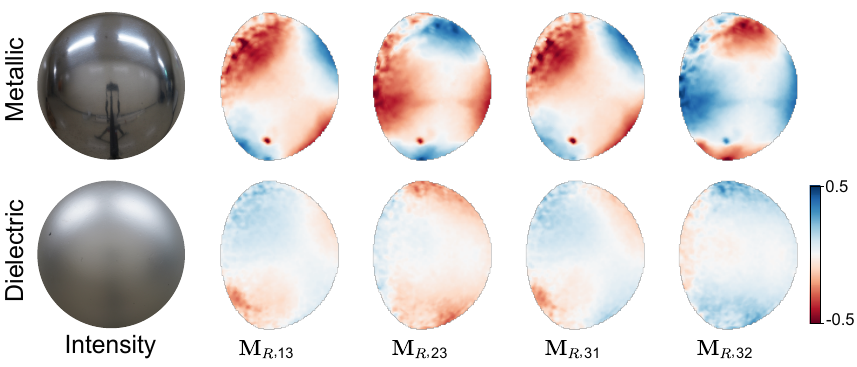}
    \vspace{-5mm}
    \caption{
    \textbf{Metallic vs. dielectric.} The retardance matrices ($\mathbf{M}_R$) of the metallic and dielectric materials at 710\,nm. 
    In their $\mathbf{M}_{R, 13}$, $\mathbf{M}_{R, 23}$, $\mathbf{M}_{R, 31}$, and $\mathbf{M}_{R, 32}$ elements, the metallic material shows pattern with opposite sign compared to the dielectric material.
    }
    \label{fig:dielec_metal}
    \vspace{-5mm}
    \end{minipage}
\end{figure*}

\paragraph{Principal Component Analysis}
Inspired by previous works~\cite{nielsen2015optimal,matusik2003data,ngan2006image,baek2021polarimetric}, we analyze low-rank structures of hpBRDF using PCA over spectral-polarimetric-angular domains.
Figure~\ref{fig:pca_main}(a) shows the dimensionality of the normalized hpBRDF. 
We also visualize the first two principal components across selected angular and spectral slices, highlighting characteristic patterns in the data in Figure~\ref{fig:pca_main}(b). These visualizations reveal distinct structural patterns within each slice, reflecting how the data varies across angular and spectral dimensions. Notably, the slices along the angular dimensions, particularly $(\theta_d, \theta_h)$, exhibit stronger and more complex variations, indicating a higher degree of angular dependency. In contrast, the variations in the spectral domain $(\theta_h, \lambda)$ appear more gradual and less spatially structured, showing that the spectral properties are relatively smoother compared to the angular variations.

\paragraph{Analytical Models}
We evaluate an analytical parametric pBRDF model~\cite{hwang2022sparse} on our hpBRDF data. Figure~\ref{fig:analytic_pbrdf} shows the fitted result on the yellow rough plastic material. While the rendered sRGB intensity appears visually similar to our measurement, notable differences are observed in the Lu-Chipman decomposition results. In the depolarization matrix $\mathbf{M}_\Delta$, the model overestimates near the edge of the sphere. Moreover, the model fails to reconstruct the non-diagonal components in the retardance matrix $\mathbf{M}_R$, which we identified in Section~\ref{subsec:metallicity} as important indicators of material metallicity. These components are either underestimated or absent in the fitted result. 

\section{Discussion}
\label{sec:discussion}



\paragraph{Image resolution}
Since hyperspectral light field cameras spatially divide a single image sensor to capture images at multiple wavelengths, the resulting spatial resolution per wavelength is limited to 410$\times$410 pixels, which is lower than that of conventional image sensors. This resolution limits the dense angular sampling on a target sphere surface.

\paragraph{Tilted QWP}
To mitigate inter-reflections between the QWP and the LP, we slightly tilted the QWP mounted on the camera side. We carefully selected the tilt angle to minimize reflections while ensuring that no perceptible optical artifacts were introduced.

\paragraph{Inpainting}
To address the missing Mueller matrices due to inevitable geometric constraints of our imaging setup, we interpolated the missing entries by finding the smoothest values in the angular domain. Although this inpainting produces visually plausible results, it does not guarantee physical validity. To improve the overall quality of the hpBRDF table, future work could incorporate physics-informed or data-driven methods that jointly model both the angular and polarization domains. Such approaches potentially produce more accurate and physically consistent estimations of the missing Mueller matrix entries.

\section{Conclusion}
\label{sec:conclusion}
In this work, we present the first hpBRDF dataset of real-world materials, leveraging our reflectance acquisition system. This dataset addresses the longstanding gap in modeling spectral and polarimetric BRDF across the visible to NIR wavelengths. It enables rendering of spectro-polarimetric images, facilitating realistic simulation of diverse imaging scenarios.

We further analyze how hpBRDF properties vary with wavelength, polarization state, material type, and geometric conditions, and propose compact representations through PCA and implicit neural representation. We anticipate that our hpBRDF dataset will contribute to spectro-polarimetric imaging and display research and inverse rendering that exploit combined spectral and polarimetric features.

\paragraph{Limitations}
Although our acquisition setup benefits from single-shot hyperspectral imaging, it also faces limitations such as reduced spatial resolution per wavelength due to the sensor's division into multiple spectral channels. 
Also, anisotropic samples cannot be captured in our imaging system. 
To handle missing Mueller matrix entries arising from imaging-system geometry, we employ inpainting, which, although plausible, may not exactly match the physical ground truth. 
Overcoming the aforementioned challenges will be an interesting area for future work.

\begin{acks}
This work was partly supported by the National Research Foundation (NRF) grant (RS-2024-00438532, RS-2023-00211658, RS-2024-00357548), and the Institute of Information \& Communications Technology Planning \& Evaluation (IITP)-ITRC (Information Technology Research Center) grant (IITP-2025-RS-2024-00437866), funded by the Korean government (MSIT).
\end{acks}


\bibliographystyle{ACM-Reference-Format}
\bibliography{references}

\clearpage
\newpage

\clearpage
\newpage

\end{document}